\documentclass[conference]{IEEEtran}
\IEEEoverridecommandlockouts
\usepackage{cite}
\usepackage{amsmath,amssymb,amsfonts}
\usepackage{algorithmic}
\usepackage{graphicx}
\usepackage{textcomp}
\usepackage{xcolor}
\usepackage{subcaption}
\usepackage{booktabs}
\usepackage{enumitem}
\usepackage{adjustbox}
\usepackage[hidelinks]{hyperref}
\usepackage{balance}
\usepackage[inkscapelatex=false]{svg}
\usepackage{xurl}
\usepackage{tikz}

\newcommand\copyrighttext{%
  \footnotesize \textcopyright \the\year{} IEEE. Personal use of this material is permitted. Permission from IEEE must be obtained for all other uses, including reprinting/republishing this material for advertising or promotional purposes, collecting new collected works for resale or redistribution to servers or lists, or reuse of any copyrighted component of this work in other works.}

\newcommand\copyrightnotice{%
\begin{tikzpicture}[remember picture,overlay]
\node[anchor=south,yshift=10pt] at (current page.south) {\fbox{\parbox{\dimexpr0.75\textwidth-\fboxsep-\fboxrule\relax}{\copyrighttext}}};
\end{tikzpicture}%
}
\def\BibTeX{{\rm B\kern-.05em{\sc i\kern-.025em b}\kern-.08em
    T\kern-.1667em\lower.7ex\hbox{E}\kern-.125emX}}
\begin{document}

\title{Evaluating Redundancy Mitigation in Vulnerable Road User Awareness Messages for Bicycles}

\author{
\IEEEauthorblockN{Nico Ostendorf\IEEEauthorrefmark{1}\IEEEauthorrefmark{2},
Keno Garlichs\IEEEauthorrefmark{1},
Lars C. Wolf\IEEEauthorrefmark{2}
}

\IEEEauthorblockA{
\IEEEauthorrefmark{1} Corporate Research, Robert Bosch GmbH, Hildesheim, Germany \\
\IEEEauthorrefmark{2} Institute of Operating Systems and Computer Networks, Technische Universität Braunschweig, Braunschweig, Germany\\
\texttt{\{nico.ostendorf,keno.garlichs\}@de.bosch.com}, \texttt{wolf@ibr.cs.tu-bs.de}
}
}
\maketitle
\copyrightnotice

\begin{abstract}
Vehicle-to-X (V2X) communication has become crucial for enhancing road safety, especially for Vulnerable Road Users (VRUs) such as pedestrians and cyclists. 
However, the increasing number of devices communicating on the same channels will lead to significant channel load. 
To address this issue this study evaluates the effectiveness of Redundancy Mitigation (RM) for VRU Awareness Messages (VAMs), focusing specifically on cyclists.
The objective of RM is to minimize the transmission of redundant information.
We conducted a simulation study using a urban scenario with a high bicycle density based on traffic data from Hannover, Germany. 
This study assessed the impact of RM on channel load, measured by Channel Busy Ratio (CBR), and safety, measured by VRU Perception Rate (VPR) in simulation. 
To evaluate the accuracy and reliability of the RM mechanisms, we analyzed the actual differences in position, speed, and heading between the ego VRU and the VRU, which was assumed to be redundant. 
Our findings indicate that while RM can reduce channel congestion, it also leads to a decrease in the perception rate of VRUs. 
The analysis of actual differences revealed that the RM mechanism standardized by ETSI often uses outdated information, leading to significant discrepancies in position, speed, and heading, which could result in dangerous situations. 
To address these limitations, we propose an adapted RM mechanism that improves the balance between reducing channel load and maintaining VRU awareness.
The adapted approach shows a significant reduction in maximum CBR and a less significant decrease in VPR compared to the standardized RM. 
Moreover, it demonstrates better performance in the actual differences in position, speed, and heading, thereby enhancing overall safety.
Our results highlight the need for further research to optimize RM techniques and ensure they effectively enhance V2X communication without compromising the safety of VRUs.

\end{abstract}


\section{Introduction}
In recent years, Vehicle-to-X (V2X) communication has emerged as an important technology for improving road safety and efficiency. 
As the group of potential V2X users continues to expand to include not only vehicles, but also Vulnerable Road Users (VRUs) such as pedestrians and cyclists, the importance of efficient communication between all parties is becoming increasingly apparent. 
With an increasing number of devices communicating on the same channels, the issue of channel congestion is becoming a major concern for the future of V2X technology.

To address this, various methods have been developed to mitigate channel congestion.
For instance, Decentralized Congestion Control (DCC) \cite{Etsi2018DCC} is an approach that aims to equitably distribute the available resources among the different V2X users.
There are also strategies to optimize message content, such as avoiding unnecessary transmission of certain information to save channel capacity. 
One such technique is Redundancy Mitigation (RM) which aims to avoid the transmission of redundant information from different participants. 
RM is a part of the VRU Awareness Message (VAM).
Redundant in this context means that VRU A's dynamic state reported in its VAM is similar enough to the ego VRUs dynamic state such that A's VAM represents both VRUs close enough for most road safety applications.
However, its application within the context of the VAM has not been extensively researched, highlighting the need to investigate its potential to reduce channel congestion without losing essential information about VRUs. 
Our focus is to evaluate RM for cyclists using metrics such as Channel Busy Ratio (CBR) and VRU Perception Rate (VPR), and to address the challenges of the standardized RM by introducing an optimized RM mechanism.

Our research specifically targets cyclists as a significant subset of VRUs, given their substantial presence and frequent involvement in road fatalities. 
Due to their driving behavior and dynamics, cyclists are among the most at-risk VRUs. 
In 2022 more than 2,000 cyclists were killed due to accidents in the EU.
Furthermore, they represent the only road user group with stagnant and not decreasing fatalities in recent years \cite{Directorate2024fatalities}.
In conjunction with the European Commission’s Vision Zero \cite{eu2020next} target, it demonstrates the critical need to address their safety and communication needs within the V2X framework. 

To accurately assess V2X communication performance, we conducted a simulation study replicating a high volume of bicycles in an urban setting.
To achieve this, we used the Hannover Traffic Scenario (HaTS) \cite{Ostendorf2025HaTS} incorporating cyclists based on traffic counts in Hannover, Germany. 
Additionally, to ensure realistic bicycle behavior, we integrated our realistic bicycle dynamics
model for SUMO \cite{Ostendorf2025Enhancing}. 
This approach allowed us to accurately assess V2X communication performance in a context reflective of real-world conditions.

The remainder of this paper is structured as follows: 
Section \ref{sec::relatedWork} provides an overview about related work. 
The fundamentals of VRUs and the standardized RM mechanism are explained in Section \ref{sec::VRU}. 
Current challenges of the standardized RM mechanism and possible solutions are explained in Section \ref{sec::ChallengesSolutions}.
Section \ref{sec::AdaptedApproach} introduces an optimized RM approach for VRUs. 
The used simulation environment is explained in Section \ref{sec::eval} as well as the results of the evaluations. 
Section \ref{sec::Conc} presents the conclusion.

\section{Related Work}
\label{sec::relatedWork}
The existing literature has a notable gap in publications that specifically address RM within the context of VAMs or VRUs. 
While there are publications that delve into RM in the context of V2X communication, particularly in the context of Collective Perception Messages (CPM), there is a lack of dedicated research on RM for VAMs.

The publication by Huang et al. \cite{Huang2020Data} examined the redundancy issue within the CPM, demonstrating the impact of redundant messages on channel load in a highway context. 
They proposed a probabilistic data selection scheme for redundant objects in the CPM, resulting in a 60\% reduction in communication overhead. 

Similarly, other publications \cite{Malik2024DistanceBased, Gokulnath2020Redundancy, Imed2023Distributed, Sakr2024Evaluation, Delooz2022Analysis}, introduced various approaches and improvements for RM, all aiming to reduce channel load and the number of redundant objects in CPMs.
%Similarly, other publications \cite{Malik2024DistanceBased, Wellington2023Redundancy, Gokulnath2020Redundancy, Imed2023Distributed, Sakr2024Evaluation, Delooz2022Analysis}, introduced various approaches and improvements for redundancy mitigation techniques, all aiming to reduce channel load and the number of redundant objects in CPMs.
Those results are not directly applicable to VAMs, as in the CPM, only redundant objects otherwise included in a generated message will be omitted, whereas in the VAM, a VRU will skip the entire message if redundancy is detected. 
Our paper addresses distinct challenges and techniques within the VAM context, setting it apart from existing research. 
This unique focus underscores the novel contributions of our work compared to previous studies.

\section{Redundancy Mitigation for VAM}
\label{sec::VRU}
\subsection{Definition of VRU and VAM}
%According to the ETSI standard \cite{Etsi2023VAM}, VRUs include the following road users: 

%\begin{enumerate}
%    \item \textit{Pedestrian}: ordinary pedestrian, road worker, first responder
%    \item \textit{Bicyclist}: bicyclist, wheelchair user, horse and rider, rollerskater, e-scooter, pedelec, e-bike
%    \item \textit{Motorcyclist}: moped, motorcycle
%    \item \textit{Animal}
%\end{enumerate}
According to the ETSI standard \cite{Etsi2023VAM}, VRUs encompass a diverse range of road users, including pedestrians, bicyclists, motorcyclists, and animals.

The standardized V2X message for VRUs is the VAM.
It provides detailed information about the current state of the VRU. 
It consists of up to six different containers, each containing specific information. 
Among these, the Basic Container and the High Frequency Container are the most crucial, as they include details about the VRUs current position, speed, heading, and other dynamic information. 
All other containers are optional, such as the Low Frequency Container.
The Low Frequency Container contains information about the VRU type and size. 
Additionally, the Cluster Operation Container and the Cluster Information Container can be utilized to cluster a group of VRUs, reducing the number of individual VAMs on the medium. 
%Clusters should handle VRUs that exhibit similar behavior and can be consolidated into one large moving VRU. 
%The Cluster Operation container is used for operations like joining or leaving a cluster, while the Cluster Information Container contains information about the cluster itself. 
Lastly, the Motion Prediction Container, includes data such as path history and path prediction.

A VAM is sent at intervals specified by [$T_{GenVamMin}$,$T_{GenVamMax}$] = [100,5000]\,ms.
The specific time between two consecutive messages depends on the rules governing the message generation process. 
A new VAM must be generated if any of the following rules apply:

\begin{enumerate}[label=(\roman*)]
\item time since last ego VAM exceeds $T_{GenVamMax}$
\item heading since last ego VAM has a minimum absolute difference of $\Delta_{\psi}$ = 4\,°
\item speed since last ego VAM has a minimum absolute difference of $\Delta_v$ = 0.5\,m/s
\item distance from last position in ego VAM has a minimum absolute difference of $\Delta_d$ = 4\,m
\item trajectory interception probability since last ego VAM has a minimum absolute difference of $\Delta_{t}$ = 10\%
\item the VRU decides to join/leave a cluster
\item the VRU detects a new vehicle coming closer than the safe distance:
\begin{itemize}
    \item Safe Lateral Distance: Max[2,A], A = distance VRU can travel lateral in $T_{GenVamMax}$\,ms
    \item Safe Longitudinal Distance: Distance ego vehicle can travel longitudinal in $T_{GenVamMax}$\,ms
    \item Safe Vertical Distance: 5\,m
\end{itemize}
\end{enumerate}
These rules should make sure that a VRU only sends a new message if there is a critical change in the ego data or if the last ego VAM was too long ago. 

\subsection{Redundancy Mitigation Rules and Process}
\label{sec::RMRules}
In addition to the message generation rules, there are also redundancy mitigation rules in place. 
These rules are designed to prevent a VRU from transmitting information about itself if another VRU has already reported redundant information. 
If the RM rules are met, a VRU is not required to send its own VAM, even if a message generation rule is met. 
A new VAM should not be generated if all of the following rules are met:

\begin{enumerate}[label=(\roman*)]
\item The time elapsed since the last ego VAM does not exceed $numSkipVamsForRedundancyMitigation \cdot T_{GenVamMax}$ = $[2,10] \cdot 5000\,ms$.
\item A VAM of reference VRU has a heading difference $<$\,4\,°,
\item an absolute speed difference $<$\,0.5\,m/s and 
\item a distance difference $<$\,4\,m to ego VRU.
\end{enumerate}

The full process of checking the redundancy between two messages is described as follows in ETSI TS 103 300-3 \cite{Etsi2023VAM}.
If the ego VRU receives a VAM it should save the information of this VAM in a local dynamic map (LDM).
Every time the ego VRU fulfills a message generation rule, this VRU should check if inside its LDM there is a received VAM fulfilling the RM rules. 
If there is such a VAM from another VRU, then the ego VRU should not transmit its own VAM, because there is already a VAM which reports redundant information. 
If there is no such VAM, then the ego VRU should send its own VAM.

\section{Current Challenges and Proposed Solutions}
\label{sec::ChallengesSolutions}

The VAM standard has some challenges that need to be addressed for better RM. 
The standard provides a procedure for comparing received VAMs with the ego status to eliminate redundancy, but it does not specify which VAMs should be used for comparison. 
This lack of detail leads to issues, especially when comparing current ego data with old messages.

One of the main problems is that there is no rule about how old a message can be for it to be used in redundancy checks. 
This means that very old messages can still be used for redundancy checks.
Using old data means that road users might not be accurately aware of their surroundings because some VRUs rely on potentially obsolete messages.

To solve this, we introduce a rule to determine if a message is outdated.
This rule calculates the expected time for receiving a new VAM from a VRU, based on the following formula:
$$t_{expected} = min(\frac{4\,m}{v - 0.5\,m/s},T_{GenVamMax}) + t_{now}$$
The latest time to expect a new VAM is when a VRU has traveled 4\,m since its last VAM or when $T_{GenVamMax}$ exceeds. 
To calculate the time it takes for a VRU to travel 4\,m, the transmitted speed of the VAM can be used.
Because of the rule, that a VRU can change its speed by less then 0.5\,m/s without sending a new message, we assume the lowest possible speed the VRU could drive without sending a new message: $v - 0.5\,m/s$.
If a new VAM is not received by $t_{expected}$, the old data should no longer be used for RM as it does not accurately represents the current status of the VRU anymore.
For all other generation rules (Section \ref{sec::VRU}), except the distance rule, it is not possible to calculate the latest time at which this rule will lead to a new VAM. 

Despite this optimization, timing issues still persist in the RM process.
When comparing ego data with the reference VAM, the reference data might be up to 5\,s old. 
As a result, the position, speed, and heading in the reference VAM could be outdated, leading to inaccurate redundancy checks. 
To address this, it could be a solution to use trajectory predictions to compare the ego data with the expected values in the reference data. 
But in this case all VRUs would need to send accurate trajectory predictions for accurate RM. 

A more effective approach would be to check redundancy at the moment the ego VRU receives a VAM, eliminating timing issues and ensuring timely and accurate redundancy checks.

\section{Adapted Redundancy Mitigation Approach}
\label{sec::AdaptedApproach}
\begin{figure}[ht]
\begin{center}
    \includegraphics[width=0.7\linewidth]{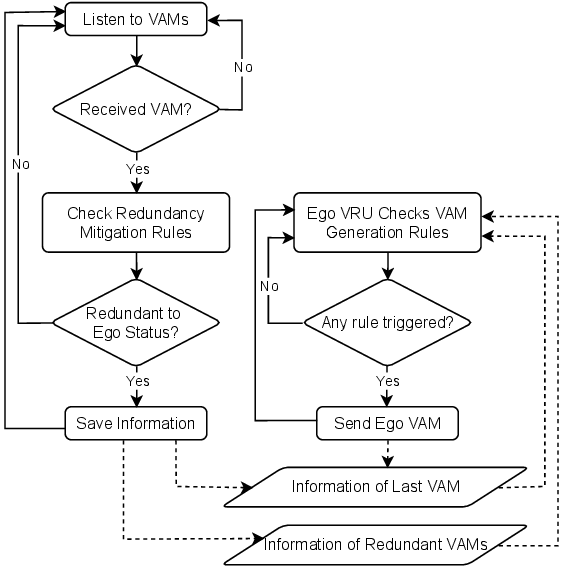}    
\end{center}
\caption{Flowchart of modified Redundancy Mitigation mechanism.}
\label{fig:flowchart}
\end{figure}

In response to the challenges faced in the standardized RM mechanism, we decided to optimize this mechanism with our previously presented solutions.
Figure \ref{fig:flowchart} illustrates a simplified flowchart of our modified RM mechanism.

The primary issue with the standardized RM was timing, particularly in determining which messages an ego VRU should use for redundancy comparison.
To address this, we have relocated the RM check to the VAM listening phase.
Once a VRU receives a VAM, the ego VRU immediately checks if the received VAM is redundant to its current ego status.
If the message is not redundant based on the RM rules (Subsection \ref{sec::RMRules}), no action is taken.
If the message is redundant to the ego status, the ego VRU saves its ego status as if it had sent an own VAM and also stores the received VAM in a list of redundant VAMs.

Additionally the process of checking for the generation of a new VAM has been modified.
As before, the ego VRU checks if the generation rules are fulfilled based on the current ego data and the data from the last ego VAM.
It is important to note that the information from the last ego VAM may be influenced by the mitigation process, as previously explained.
If the generation rules are fulfilled, the ego VRU generates a new VAM and clears the list of redundant VAMs. 
In cases where the generation rules are not fulfilled, the ego VRU conducts an additional check with the redundant VAMs list.
If the list is empty, no further checks are performed.
However, if the list contains VAMs, the ego VRU checks for each VAM in the list if the generation rules are fulfilled, if it compares its current ego status with the information in that VAM.
Any VAM fulfilling the generation rules is then removed from the redundant VAMs list.
If the list is empty after this process, the ego VRU will generate a new ego VAM. 

The new mechanism resolves the timing problem, by conducting the redundancy check always at the moment a new VAM is received. 
The additional generation rule check ensures that the ego VRU remains in range of at least one redundant object.

\section{Evaluation}
\label{sec::eval}

The effectiveness of the standardized RM and of the proposed RM mechanism for the VAM as well as the potential influence on safety, is evaluated through an extensive simulation study.
The following subsections provide a detailed overview of the employed simulation environment, the simulation scenario, parameter variations, and present a comprehensive analysis of the simulation results.

\subsection{Simulation Environment}
Table \ref{tab:SimParams} details the important simulation parameters employed in this study. 

\begin{table}[ht!]
\centering
\caption{Simulation Parameters}
\begin{tabular}{ll}
\toprule
\textbf{Parameter} & \textbf{Value}\\\midrule
Protocol Stack & ITS-G5 \\
Frequency Band & 5.9\,GHz \\
Bitrate & 6\,Mbps \\
Obstacle Loss & Dielectric Obstacle Loss \\
Transmission Power & 23\,dBm \\
Scenario & HaTS \cite{Ostendorf2025HaTS} \\
Seed Amount & 4 \\
Simulation Start Time & 7250\,s \\
Simulation Period & 50\,s (incl. 10\,s of warmup) \\\bottomrule
\end{tabular}
\label{tab:SimParams}
\end{table}

All simulations have been performed on a ITS-G5-based V2X network.
To realize this network, we used the discrete event simulator OMNeT++ \cite{Varga2010Omnet} with its framework INET in combination with the vehicular network simulation framework Veins\cite{Sommer2011Bidirectionally}. 
To accurately model node mobility, OMNeT++ was coupled with the microscopic traffic simulator SUMO \cite{Lopez2018Microscpic}.

Given that these frameworks do not natively support the VAM, a new message type with associated triggering conditions has been implemented according to the specifications \cite{Etsi2023VAM}. 
We implemented the VAM with a static message size of 300 Bytes. 
This choice was made to strike a balance between the assumptions made by the CAR 2 CAR Communication Consortium's spectrum needs analysis \cite{c2c2020Road}, which specifies an average VAM size of around 235 Bytes, and the 5G Automotive Association's spectrum needs analysis \cite{5gaa2021Study}, which specifies an average VAM size of 350 Bytes.

\begin{figure}[ht]
\begin{center}
    \includegraphics[width=0.7\linewidth]{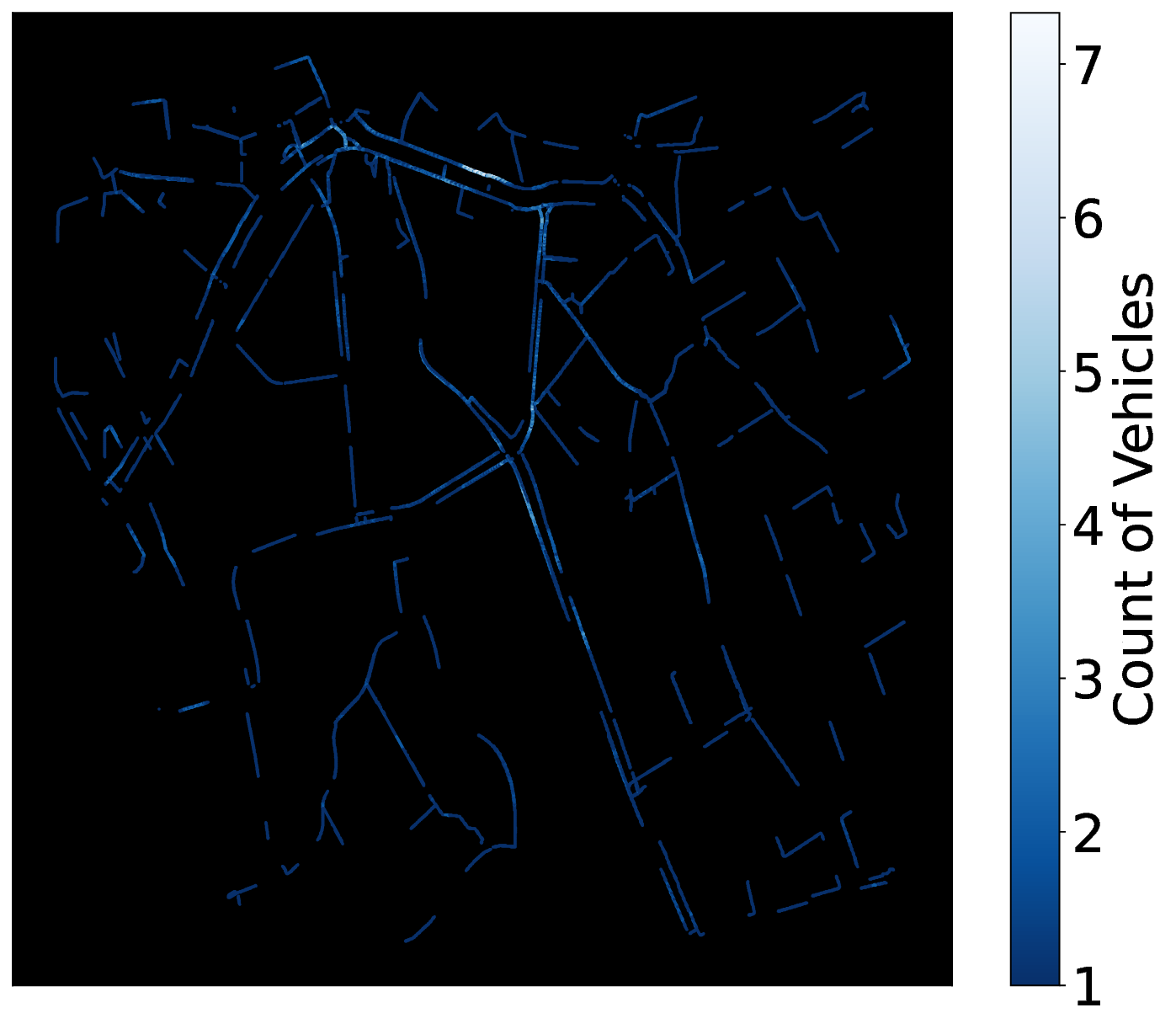}    
\end{center}
\caption{Map of bicycle density of HaTS.}
\label{fig:density}
\end{figure}

\begin{figure}[ht]
\begin{center}
    \includegraphics[width=0.7\linewidth]{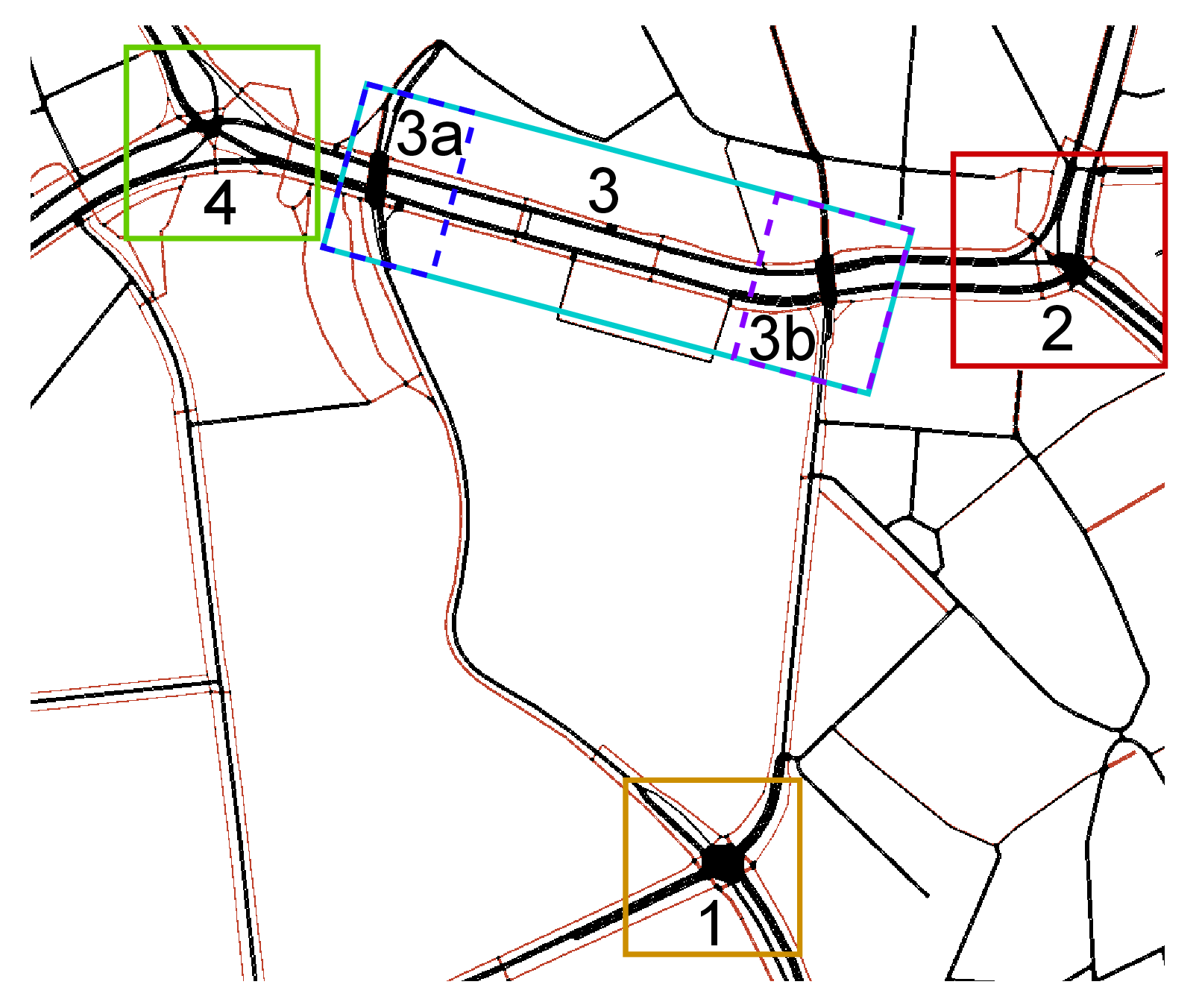}    
\end{center}
\caption{Regions of Interest in HaTS.}
\label{fig:positions}
\end{figure}

To enhance the realism of bicycle dynamics within SUMO, we integrated our open-source Realistic Bicycle Dynamics Model\footnote{\url{https://github.com/boschresearch/RealisticBicycleDynamicsModel}} \cite{Ostendorf2025Enhancing}. 
As simulation scenario we selected our open-source Hannover Traffic Scenario (HaTS)\footnote{\url{https://github.com/boschresearch/HanoverTrafficScenario}} \cite{Ostendorf2025HaTS}, which is well suited for this study due to its realistic and high-density bicycle traffic representation.
Figures \ref{fig:density} and \ref{fig:positions} provide an overview of the scenario.

The simulation start time was set to 7250\,s, corresponding to the peak traffic period between 8:00 AM and 8:15 AM in HaTS. 
Each simulation run spanned 50\,s, including a 10\,s warm-up period to mitigate the impact of initial conditions and more accurately reflect real-world V2X participation initiation.
Figure \ref{fig:density} illustrates the bicycle density distribution during the simulation period.

For the evaluation, we decided to divide the scenario into four different Regions of Interest (ROI).
These ROIs were chosen based on their high bicycle density and intersection presence, thus maximising the potential for RM events. 
Figure \ref{fig:positions} shows all the ROIs represented by rectangles.
To include other locations with lower numbers of cyclists, we also used the full scenario for evaluation.

To ensure the statistical robustness of the results and mitigate the influence of specific initial conditions, each configuration were simulated by four distinct random seeds. 
This approach enhances the reliability and generalizability of the findings.

\subsection{Metrics}

For the evaluation we selected two important metrics to study the effectiveness on the channel load and the impact for safety. 
For the channel load we used the Channel Busy Ratio (CBR) as defined in ETSI EN 302 571 \cite{Etsi2017Radio}: 
$$CBR = \frac{t_{busy}}{t_{CBR}}$$

For the safety aspect we use the VRU Perception Rate (VPR). 
The VPR represents the amount of VRUs the ego vehicle is aware of compared to all VRUs in a specific range: $$VPR=\frac{\sum x_{a,t}}{\sum x_{r,t}}$$
where $x_{a,t}$ describes the VRUs an ego vehicle is aware of at time $t$ and $x_{r,t}$ describes the amount of VRUs in a radius $r$ around an ego vehicle at time $t$.
In our evaluation, we utilized a 50\,m radius as this represents an area where awareness of other road participants is crucial, as the likelihood of interaction with them is much higher compared to VRUs at a further distance.

\subsection{Results}

\begin{figure*}[ht]
    \begin{minipage}[b]{0.22\textwidth}
        \includegraphics[width=\textwidth]{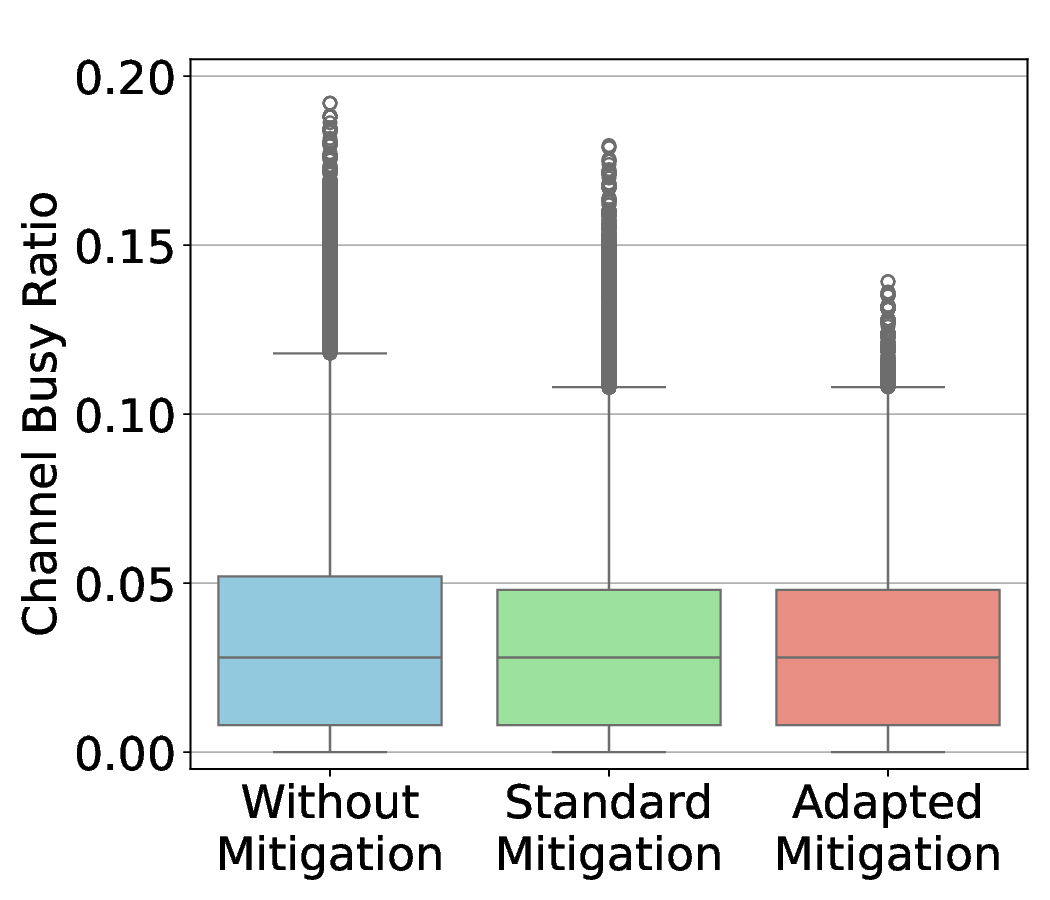}
        \subcaption{Overall}
        \label{fig:CBR_full}
    \end{minipage}
    \hfill
    \begin{minipage}[b]{0.22\textwidth}
        \includegraphics[width=\textwidth]{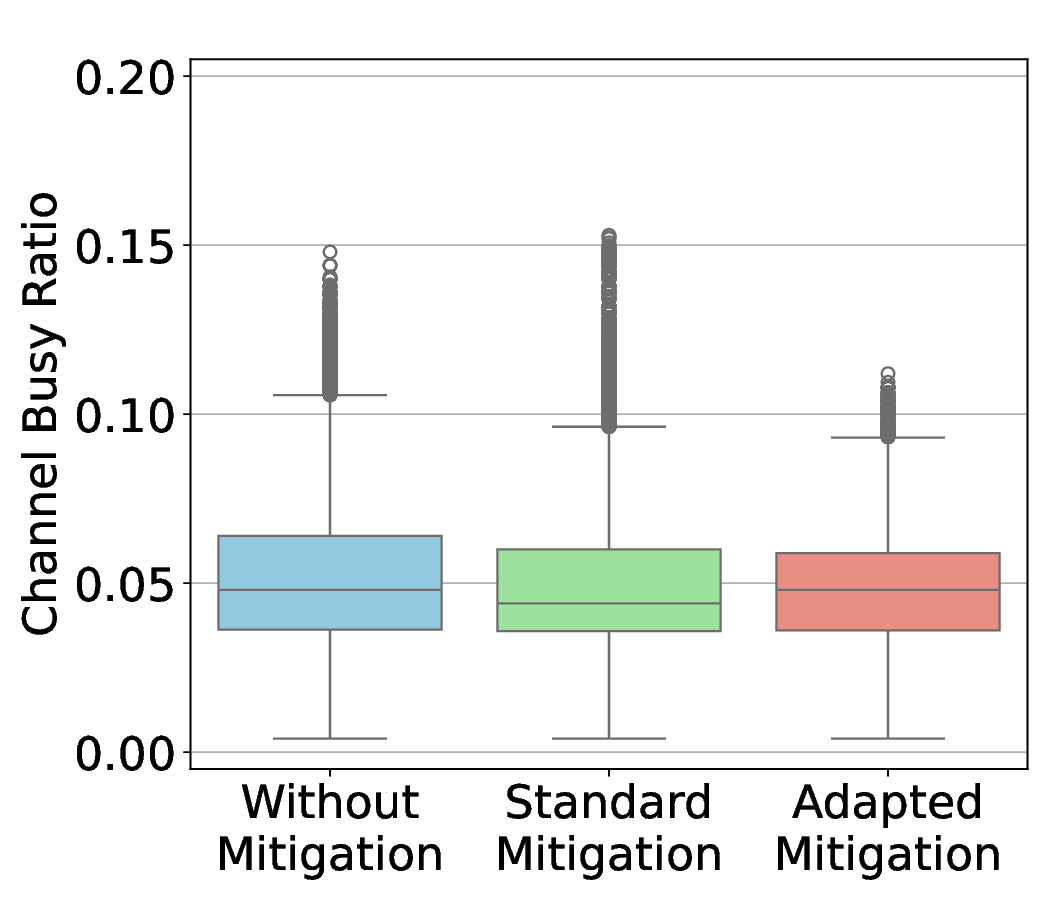}
        \subcaption{Position 1}
        \label{fig:CBR_position1}
    \end{minipage}
    \hfill
    \begin{minipage}[b]{0.22\textwidth}
        \includegraphics[width=\textwidth]{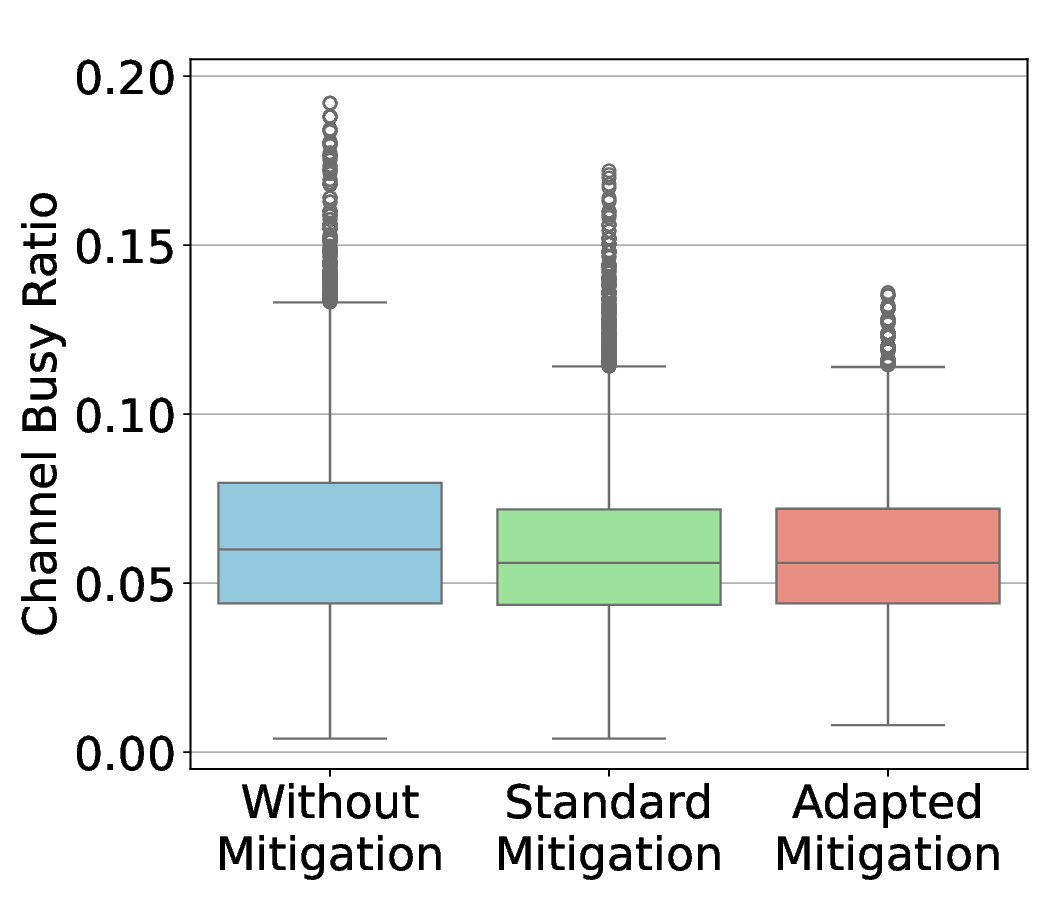}
        \subcaption{Position 3a}
        \label{fig:CBR_position3a}
    \end{minipage}
    \hfill
    \begin{minipage}[b]{0.22\textwidth}
        \includegraphics[width=\textwidth]{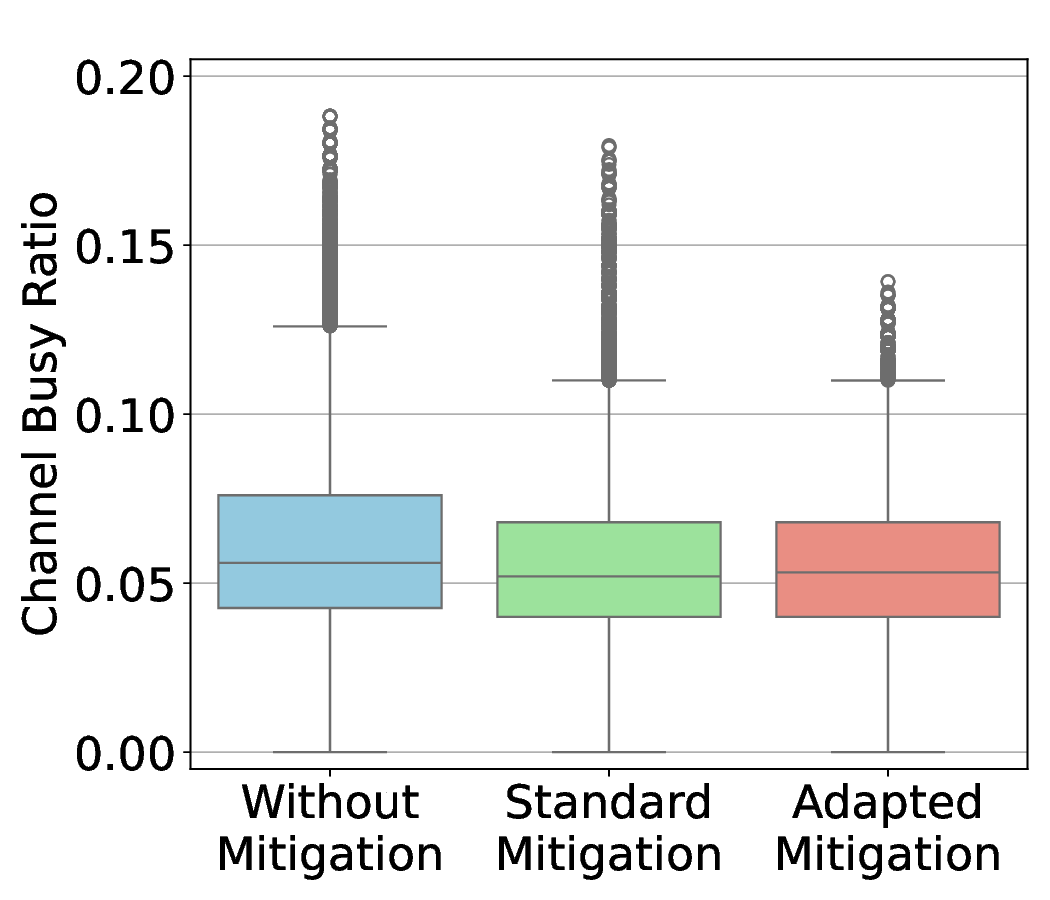}
        \subcaption{Position 4}
        \label{fig:CBR_position4}
    \end{minipage}
    \caption{Channel busy Ratio with different Redundancy Mitigation Rules at different Positions.}
    \label{fig:CBR}
\end{figure*}

The results of the evaluation of the CBR are plotted in Figure \ref{fig:CBR} as boxplots.
The median is shown by the black line. 
The upper and lower whisker of the boxplots are calculated as 1.5 times the interquartile range (IQR), with the maximum or minimum value being chosen if it is smaller than 1.5 times IQR. 
We compared the simulated HaTS without RM, with the standardized RM and with our adapted RM approach.
The standardized RM exclusively utilizes VAMs that are newer than the computed $t_{expected}$ for the specific message, ensuring a fair and equitable comparison among the different approaches.

The results presented in Figure \ref{fig:CBR_full} illustrate the impact on the CBR for the entire scenario.
Notably, the median CBR impact due to bicycles across the scenario is low, with 2.8\%. 
Consequently, the impact of RM in median is also limited considering the entire scenario.
The median CBR remains unchanged across all RM configurations, with the 25th and 75th percentiles exhibiting negligible variations.
However, the maximum CBR values reveal a significant impact of RM. 
By ignoring RM the maximum CBR in the scenario is up to 19.2\%. 
The standard RM approach offers a marginal reduction to 18\%, whereas the adapted RM significantly lowers the maximum CBR to 13.9\%.  
The low impact on the median CBR can be attributed to the low bicycle density in certain areas of the scenario.
This prompts further investigation into regions with higher bicycle concentrations (c.f. Figure \ref{fig:positions}).

Figure \ref{fig:CBR_position1} details the CBR results for position 1.
In this case the impact on the CBR is already higher compared to the overall view due to the bicycles. 
Nevertheless, the median CBR remains relatively consistent across all cases: 4.8\% without mitigation and with the adapted RM, and 4.4\% with the standard RM. 
The maximum CBR, however, mirrors the trend observed in the overall scenario. 
The maximum CBR is for the case without mitigation (14.8\%) and with standard mitigation (15.2\%) very similar.
The adapted RM significantly reduces the maximum CBR to 11.2\%.

Figures \ref{fig:CBR_position3a} and \ref{fig:CBR_position4} represent position 3a and 4.
These regions exhibit similar trends. 
The median CBR is very similar for all three cases, with a difference of less than 0.4\%.
The maximum CBR is again the highest for the case without RM, a little bit lower for the standard RM and significant lower for the adapted RM. 
Due to space constraints, results for positions 2, 3, and 3b are omitted, as they mirror the trends observed in the presented positions.

In summary, RM has the potential to decrease the CBR in high-density scenarios. 
While both approaches are not able to reduce the median CBR significantly, RM offers an advantage in mitigating peak congestion.
Our adapted RM mechanism consistently demonstrated superior performance in minimizing the maximum CBR compared to the standard RM.

\begin{figure*}[ht]
    \begin{minipage}[b]{0.21\textwidth}
        \includegraphics[width=\textwidth]{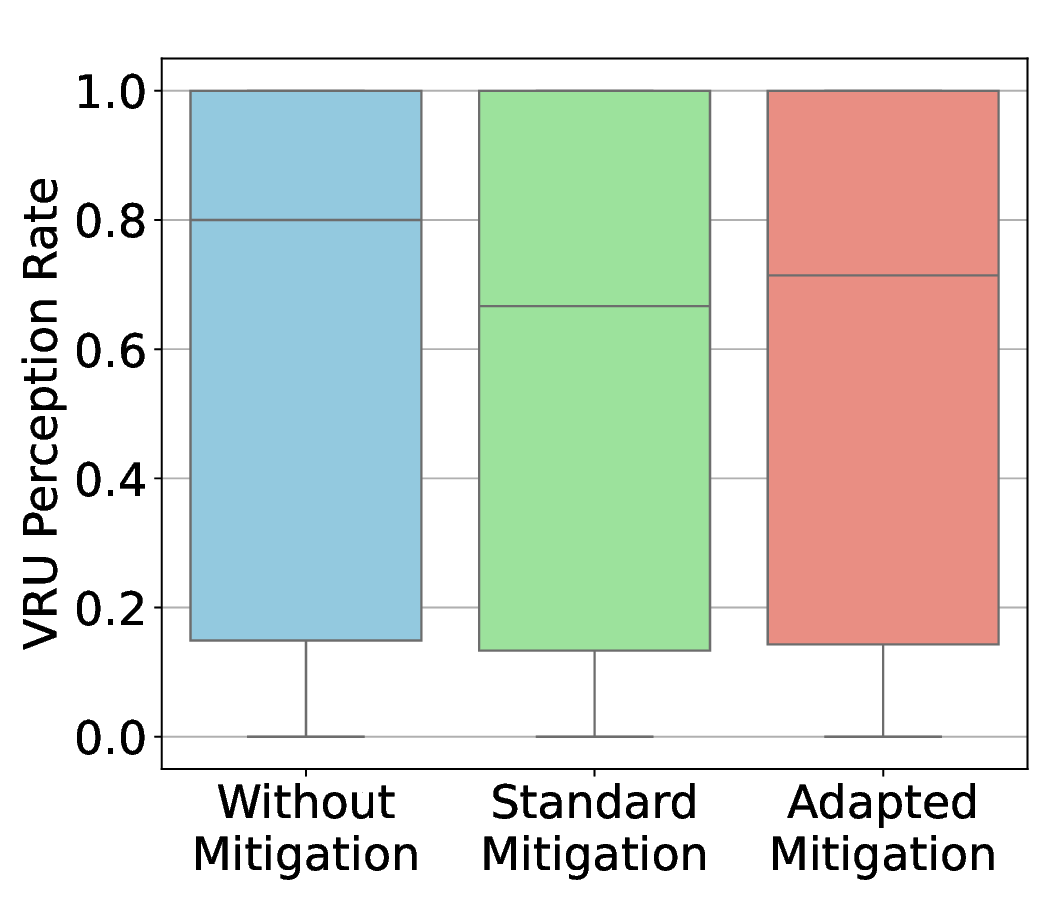}
        \subcaption{Overall}
        \label{fig:vpr_full}
    \end{minipage}
    \hfill
    \begin{minipage}[b]{0.21\textwidth}
        \includegraphics[width=\textwidth]{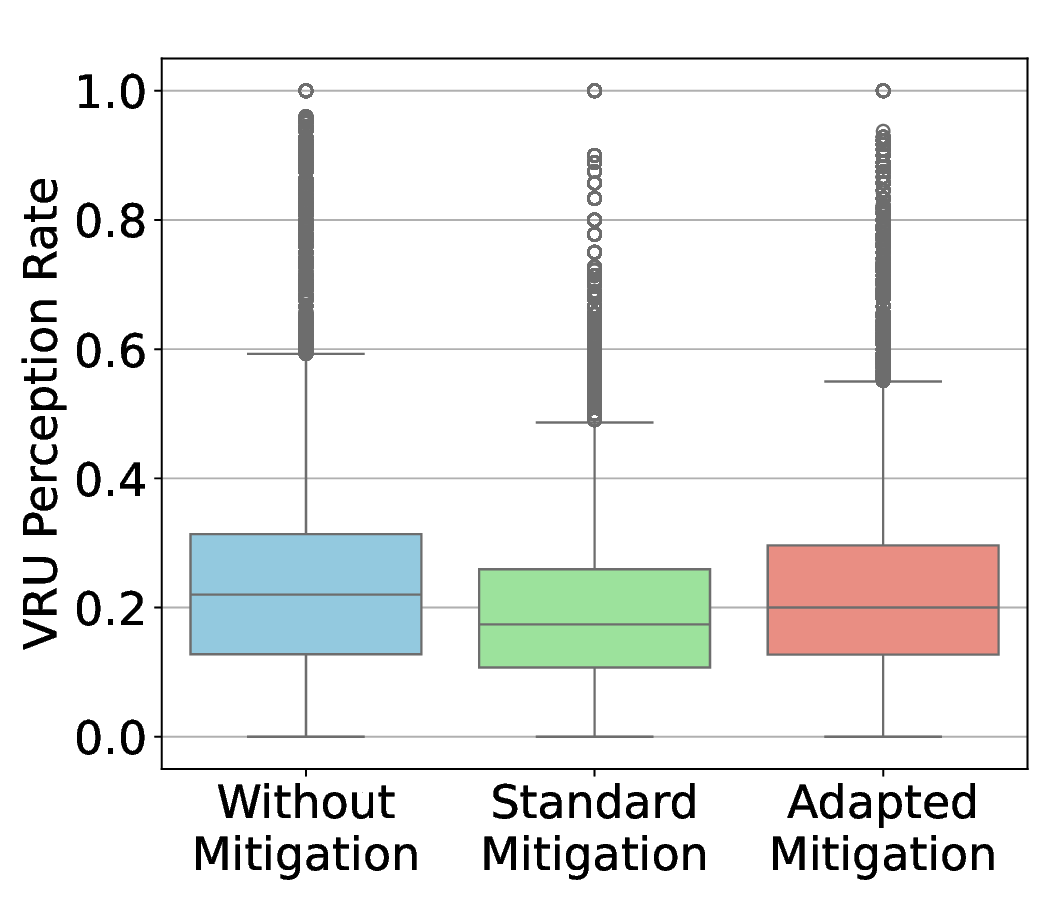}
        \subcaption{Position 1}
        \label{fig:vpr_position1}
    \end{minipage}
    \hfill
    \begin{minipage}[b]{0.21\textwidth}
        \includegraphics[width=\textwidth]{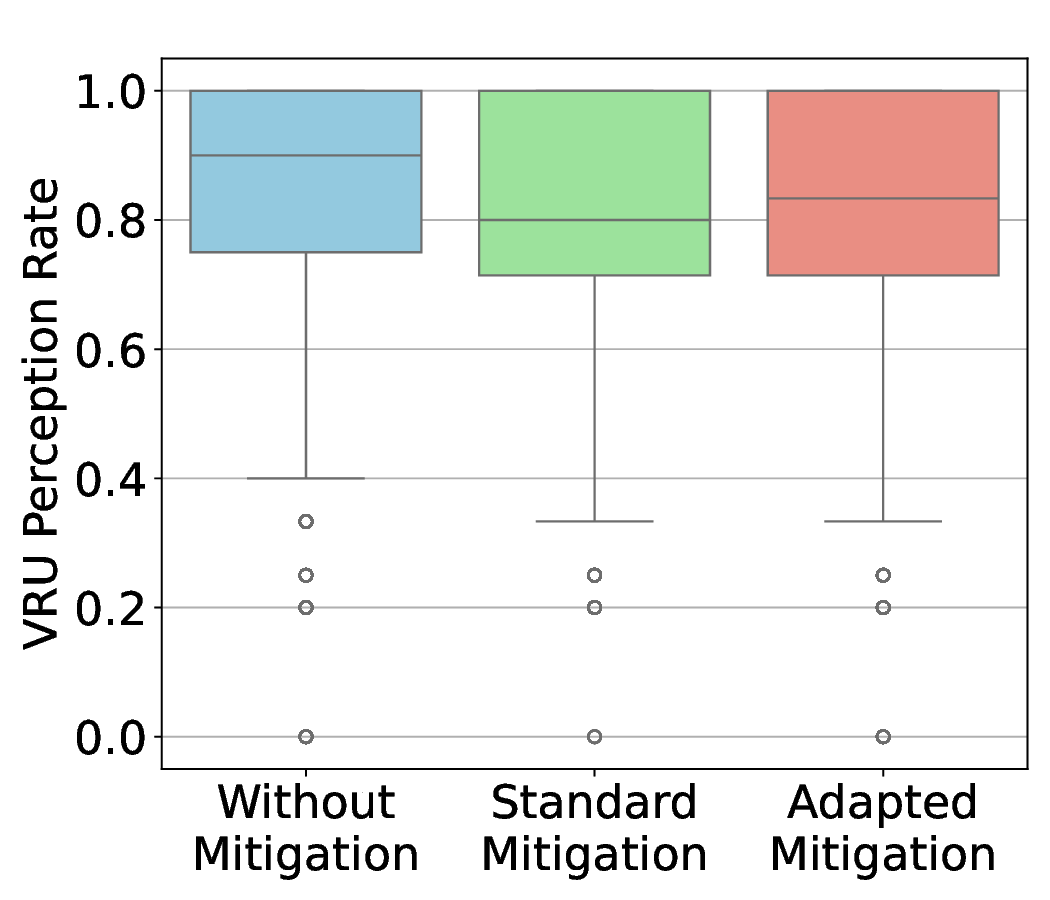}
        \subcaption{Position 3a}
        \label{fig:vpr_position3a}
    \end{minipage}
    \hfill
    \begin{minipage}[b]{0.21\textwidth}
        \includegraphics[width=\textwidth]{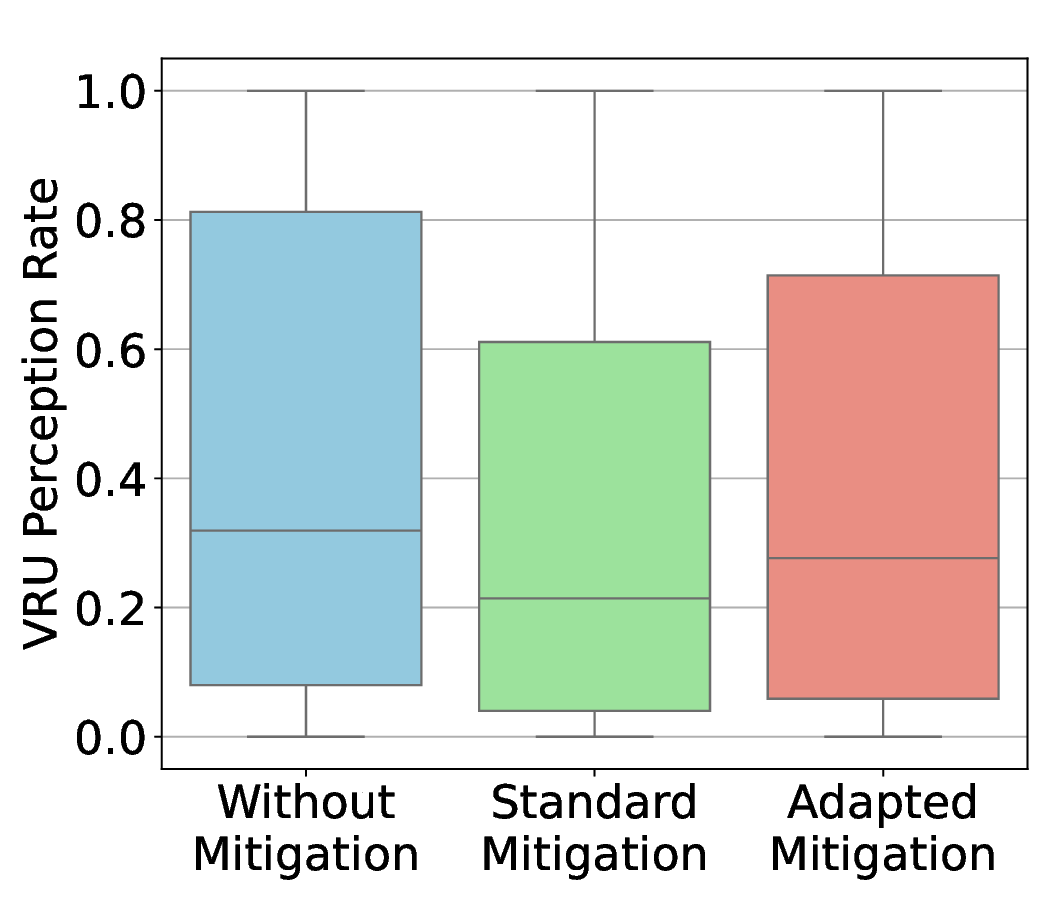}
        \subcaption{Position 4}
        \label{fig:vpr_position4}
    \end{minipage}
    \caption{VRU Perception Rate with different Redundancy Mitigation Rules at different Positions.}
    \label{fig:VPR}
\end{figure*}

For a more comprehensive analysis of the RM mechanisms, it is crucial to assess whether the reduced CBR would result in a lower perception rate of VRUs, which would reduce the overall usefulness.
A lower perception rate could lead to a critical situation in road traffic. 
The results for the VPR evaluations are represented in Figure \ref{fig:VPR}.

Figure \ref{fig:vpr_full} presents the VPR evaluation for the entire simulation scenario.
Despite the minimal impact on median CBR observed in the overall scenario, the VPR exhibits a significant reduction with the implementation of RM.
Specifically, the standard RM reduces the median VPR from 80\% (without RM) to 66.6\%, while the adapted RM reduces it to 71.4\%.
This indicates that while the adapted RM effectively minimizes maximum CBR, it results in an 8.6 percentage points reduction in median VPR. 
In contrast, the standard RM, despite its limited impact on maximum CBR, leads to a significant reduction in median VPR of 13.4 percentage points.

At position 1 (Figure \ref{fig:vpr_position1}), the VPR is notably lower compared to the overall scenario. 
The reason for this is the topology of the crossing and a big building with high signal attenuation between the two main roads. 
The standard RM again significantly reduces the median VPR to 17.4\% from 22\% (without RM), while the adapted RM results in a smaller reduction leading to a median VPR of 20\%.

Figures \ref{fig:vpr_position3a} and \ref{fig:vpr_position4} representing positions 3a and 4, while representing a very similar trend. 
The VPR is again decreased by both RM mechanisms compared to the case without mitigation.
The decrease due to standard RM is significant, while the decrease is lower and not as significant for the adapted approach.
Positions 2, 3, and 3b yielded similar results to the previous positions for the VPR.

Overall, standard RM led to a significant decrease in VPR, potentially resulting in a loss of information about the presence of up to 13.4\% of VRUs.
Although the adapted RM reduced the maximum CBR significantly more than the standard RM and mitigated the problem of high VPR losses, it still resulted in a loss of up to 8.6\% of VRUs compared to not using RM, highlighting a trade-off between CBR reduction and VRU perception.

\begin{figure*}[ht]
    \begin{minipage}[b]{0.22\textwidth}
        \includegraphics[width=\textwidth]{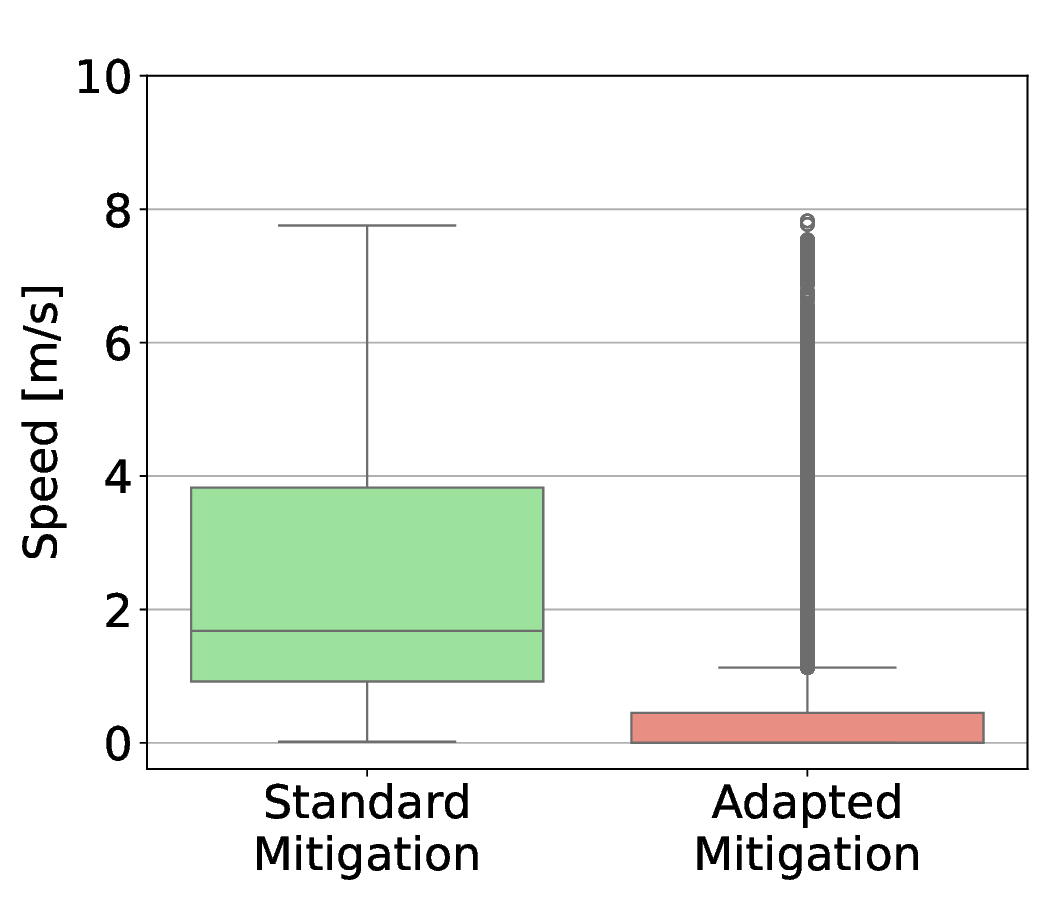}
        \subcaption{Current speeds of bicycles when detecting a redundant VAM.}
        \label{fig:speedAtMiti}
    \end{minipage}
    \hfill
    \begin{minipage}[b]{0.22\textwidth}
        \includegraphics[width=\textwidth]{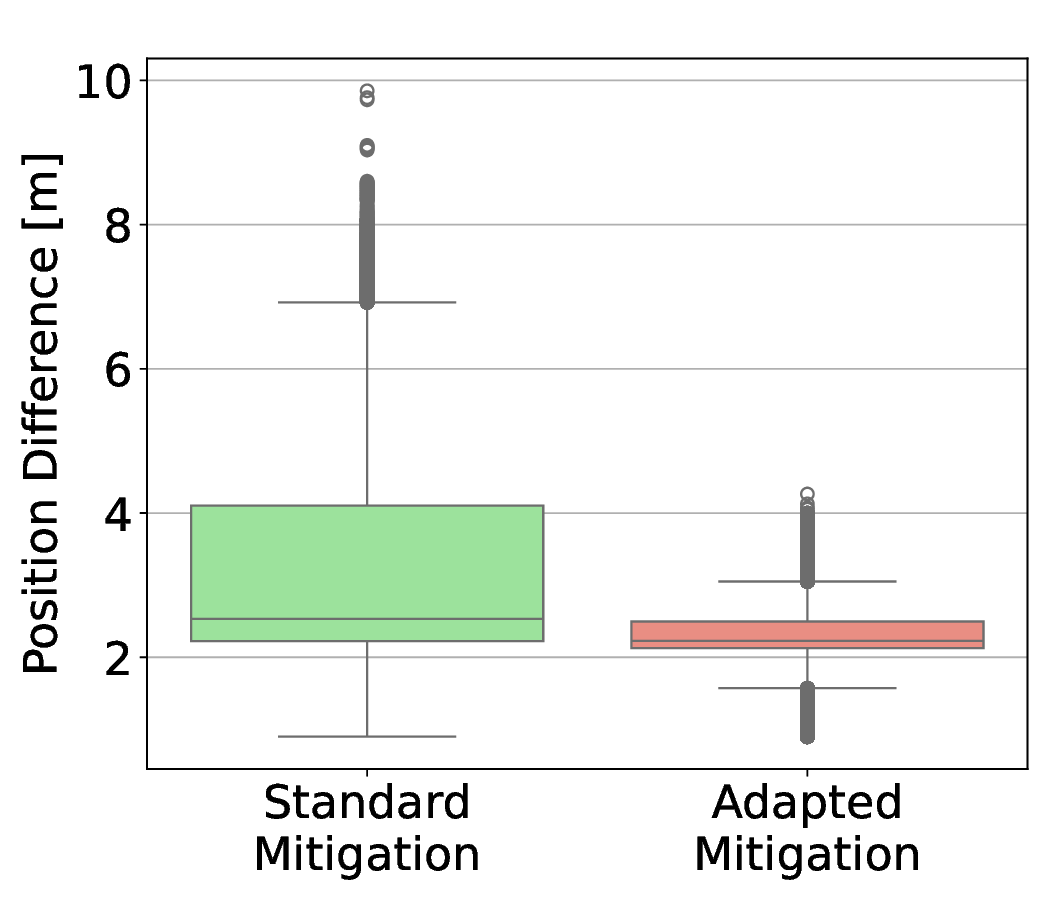}
        \subcaption{Actual position difference between the ego and redundant VRU.}
        \label{fig:posDiffAtMiti}
    \end{minipage}
    \hfill
    \begin{minipage}[b]{0.22\textwidth}
        \includegraphics[width=\textwidth]{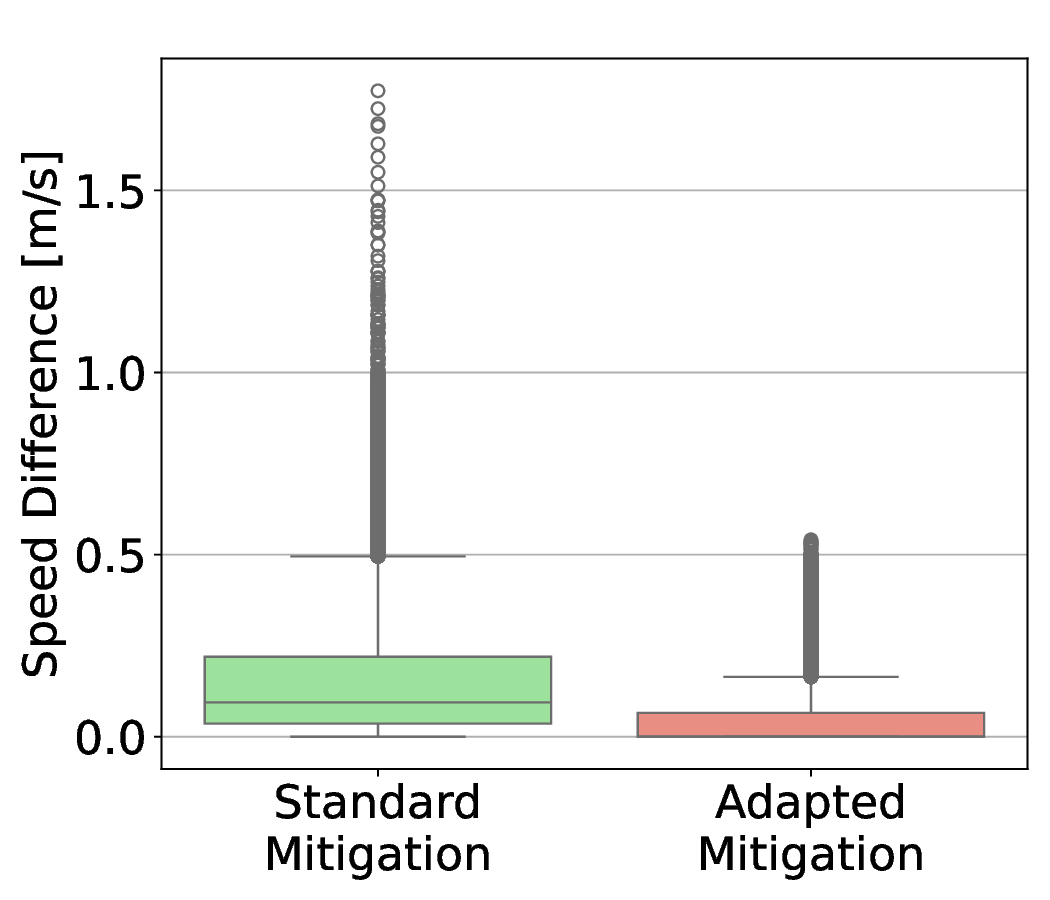}
        \subcaption{Actual speed difference between the ego and redundant VRU.}
        \label{fig:speedDiffAtMiti}
    \end{minipage}
    \hfill
    \begin{minipage}[b]{0.22\textwidth}
        \includegraphics[width=\textwidth]{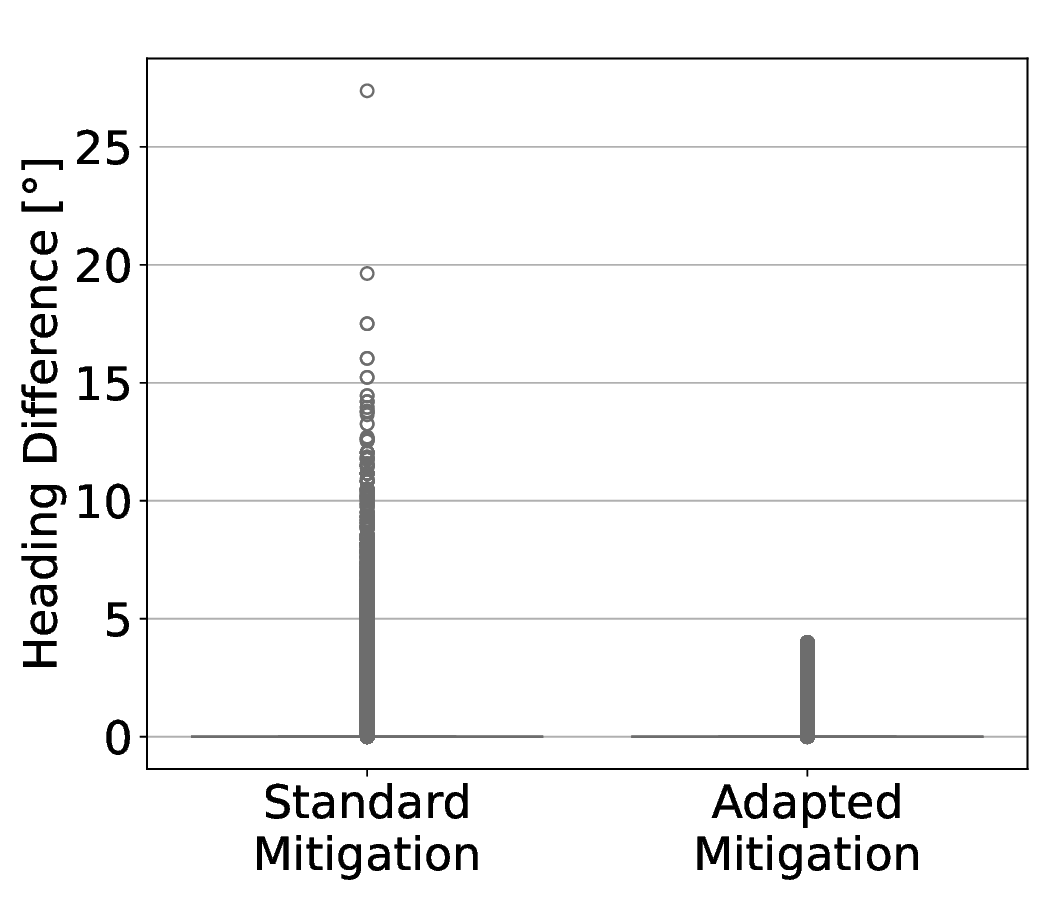}
        \subcaption{Actual heading difference between the ego and redundant VRU.}
        \label{fig:headDiffAtMiti}
    \end{minipage}
    \caption{Analysis of differences between standard and adapted redundancy mitigation mechanisms.}
    \label{fig:diffMitis}
\end{figure*}

We conducted also an analysis to examine the activation of RM mechanisms and assess whether these activations occur in critical situations.
Specifically, we analyzed the speed of a VRU when detecting a redundant VAM and suppressing the own message.
Additionally, we examined the actual position, speed, and heading difference between the ego VRU and the redundant VRU.
The results of this analysis are plotted in Figure \ref{fig:diffMitis}.

Figure \ref{fig:speedAtMiti} represents the ego speed when detecting a redundant VAM.
The standard RM exhibits a median speed value of 1.7\,m/s, with the 75th percentile at 3.8\,m/s and the 25th percentile at 0.9\,m/s, suggesting that it is primarily activated in slow driving situations.
On the other hand, the adapted RM has a median speed value of 0\,m/s, with the 75th percentile at 0.5\,m/s indicating activation mainly in standstill situations.
Given that it is potentially less critical to not send messages when the ego VRU is stationary among a group of VRUs compared to not sending while driving, the adapted RM appears to be the preferable option based on this speed evaluation.

The difference of the ego position and the actual position of the redundant VRU is plotted in Figure \ref{fig:posDiffAtMiti}.
According to the RM rules, the maximum allowable distance between the ego VRU and a redundant VRU is 4\,m. 
However, the utilization of old messages in the standardized RM can lead to distances exceeding this limit for the current actual positions.
In the case of the standard RM, the median distance is 2.5\,m, which is below the allowed 4\,m.
Nevertheless, over 25\% of all distances exceed 4\,m, with the maximum distance between an ego and redundant VRU reaching 9.9\,m.
This significant deviation from the RM rules occurs if a message from the expected redundant VRU fails to reach the ego VRU, potentially due to signal loss caused by obstructions such as buildings.
In this case, the outdated messages are used for redundancy comparison, resulting in a high distance between the VRUs.

The adapted approach has a median distance of 2.2\,m and a maximum value of 4.3\,m.
This outcome was anticipated, as the received message is directly interpreted and the distance difference between expected and actual distance is solely influenced by the sending and computing time.

The speed difference is presented in Figure \ref{fig:speedDiffAtMiti}.
The results are similar to the results for the position difference. 
The median is for both approaches in the range of the rule, but the maximum value for the standard approach is far above, with 1.8\,m/s.

The results of the heading difference (Figure \ref{fig:headDiffAtMiti}) show a similar behavior.
The maximum value of the standard mitigation is 27.4\,°. 
In comparison the maximum value of the adapted approach is 4\,°.

The results of this analysis showed clearly the problem of the standardized RM and the usage of older messages for redundancy checks. 
Also another weakness of the standardized RM could be seen.
If the ego VRU fails to receive the new VAM of another VRU due to packet loss, it is possible that the ego VRU may not only use data that is up to 5\,s old, but also outdated data due to a missed VAM, for RM purposes.
This could lead to position differences of up to 9.9\,m, heading differences of up to 27.4\,° or speed differences of 1.8\,m/s. 
Such differences could lead to very dangerous situations.
Our proposed RM shows significant improvements in the actual differences between ego VRU and redundant VRU.
It is also not vulnerable to lost or old messages.
Based on the presented results, the loss in the VPR for the adapted RM is not as critical as for the standardized RM, as the differences between the ego VRU and redundantly reported VRU are minimal. Additionally, the ego speed is low when omitting an ego VAM.

\section{Conclusion and Discussion}
\label{sec::Conc}
\balance

The analysis has revealed that RM is able to reduce the CBR in high density scenarios and especially the maxima could be significantly reduced. 
An example is the synchronized departure of multiple V2X participants from a traffic light, causing a high CBR as numerous bicycles announce their state change almost simultaneously. 
However, the inherent consequence of reducing transmissions via RM is a decrease in the VPR, which could lead to hazardous situations due to missing or outdated information about surrounding road users.

Our analysis identifies inaccuracies in the standardized RM approach, notably its lack of criteria for the age of the messages being compared. 
This poses considerable risks, as the ego vehicle may incorrectly assume it is represented by another participant. 
Consequently, the ego vehicle may suppress transmissions even when significant differences exist, such as observed distance differences of up to 9.9 meters, as well as variations in speed and heading. 
Additionally, the standardized approach shows limited effectiveness in reducing peak CBR, with a maximum reduction of 2 percentage points, and has a negligible impact on median CBR, while significantly reducing median VPR by up to 13.4 percentage points.

In contrast, the adapted RM approach presented in this paper addresses these inaccuracies by enforcing an immediate redundancy check upon message reception. 
This ensures that comparisons are always made against the most recently received data, keeping the actual differences in distance, speed, and heading between the ego and redundant road user within predefined thresholds. 
This leads to a significantly higher degree of similarity. 
Our adapted RM shows improved performance, achieving a notable reduction in maximum CBR by 5.6 percentage points, while resulting in a VPR reduction, with only an 8.6 percentage point in the median.

Further investigation into the conditions for detecting redundancy revealed another crucial difference. 
The adapted approach primarily identified redundancy when bicycles were stationary or starting to move, whereas the standardized RM more frequently triggered mitigation while bicycles were already in motion. 
From a safety perspective, suppressing updates from stationary or slowly accelerating entities is less critical than suppressing information from moving road users, further supporting the advantages of the adapted approach.

Considering these findings, the adapted RM approach significantly improves upon the current standard. 
However, the tradeoff between CBR and VPR reduction requires careful evaluation. Given the generally low median CBR in this scenario, the necessity for an RM mechanism may be questioned. 
Nonetheless, significant CBR peaks (up to 19.2\%) indicate that RM remains relevant for managing congestion. 
As noted in other studies \cite{Xhoxhi2023First}\cite{c2c2020Road}, urban environments with diverse message types (e.g., CAM and CPM) are prone to high channel loads. 
In such contexts, the adapted RM could effectively mitigate CBR peaks, and the resulting VPR reduction may be an acceptable compromise, especially since high channel loads can lead to message collisions and a further decline in VPR. 
Alternative congestion control mechanisms, such as DCC, may be less effective for short, sharp peaks due to their inherent convergence time \cite{Soto2019Strength}.
Based on the presented results the current standardized RM is not appropriate and should be replaced with a more efficient mechanism, as proposed herein. 
The proposed mechanism offers significant improvements in terms of CBR, VPR and similarity. 

\bibliographystyle{IEEEtran}
\bibliography{bibliography}
%\nocite{*}
\end{document}